\documentclass[twocolumn,superscriptaddress,showpacs,prl]{revtex4} 
\usepackage{graphicx}
\usepackage{amsmath}

\begin{document}


\title{Scalable Spin Amplification with a Gain over a Hundred}


\author{Makoto Negoro, Kenichiro Tateishi, Akinori Kagawa, Masahiro Kitagawa}
\affiliation{Division of Advanced Electronics and Optical Science,\\
Graduate School of Engineering Science, Osaka University,\\
Toyonaka, Osaka, 560-8531, Japan.}


\date{\today}

\begin{abstract}
We propose a scalable and practical implementation of spin amplification which does not require individual addressing nor a specially tailored spin network.
We have demonstrated a gain of 140 in a solid-state nuclear spin system of which the spin polarization has been increased to 0.12 using dynamic nuclear polarization with photoexcited triplet electron spins.
Spin amplification scalable to a higher gain opens the door to the single spin measurement for a readout of quantum computers as well as practical applications of nuclear magnetic resonance (NMR) spectroscopy to infinitesimal samples which have been concealed by thermal noise.
\end{abstract}


\maketitle


A magnetic moment of a single nuclear spin is so small that its induction signal is buried under thermal noise.
The minimum detectable number of perfectly polarized nuclear spins is still in the millions with a cryogenic inductive detection circuit.
If information of a single nuclear spin is quantum-logically copied to a large number of spins, a spin component can be amplified and hence its signal may be detected with sufficient signal-to-noise ratio (SNR).
This scheme is called {\it spin amplification}, which can be realized by the quantum circuit shown in Fig.~1(a)~\cite{DiVincenzo00}.
Although copy of an arbitrary, unknown state itself is prohibited by the no-cloning theorem~\cite{Wooters82}, the information about whether the state is in $|0\rangle$ or $|1\rangle$ can be copied.
Fig.~1(a) can be used for a readout of quantum computers.

We propose other potential applications of spin amplification.
To experimentally determine an arbitrary, unknown quantum state $|\psi\rangle$ of a spin, quantum state tomography is used, in which the state $|\psi\rangle$ is repeatedly prepared and measured~\cite{NielsenChuang}.
If the number of repetitions is $N$ and the measurement is limited by the thermal noise of the detection apparatus, then the SNR is increased by a factor of $\sqrt{N}$.
If the state is prepared and accumulated elsewhere to make $N$ identical copies of $|\psi\rangle$, as shown in Fig.~1(b), and measured at once, then the signal becomes $N$ times as large and the SNR is increased by a factor of $N$, which is non-trivial because it is much more efficient than $\sqrt{N}$ attained by the simple repetition.
This scheme can also be regarded as amplification in a broader sense because it improves the SNR against the noise entering afterwards.
The spin amplification (Fig.~1(b)) can be applied to quantum metrology which estimates a small unknown parameter $\phi$ contained in a quantum evolution $U$ from the initial state $|0\rangle$ to the final state $|\psi\rangle$~\cite{Ramsey}.
This procedure is regarded as spectroscopy as well when the parameter $\phi$ is given by an internal Hamiltonian of a spin system of interest.
The minimum detectable value of $\phi$ is improved by a factor of $N$ as long as it is limited by the thermal noise of the detection apparatus.
\begin{figure}[tbp]
\center{\includegraphics[width=\linewidth]{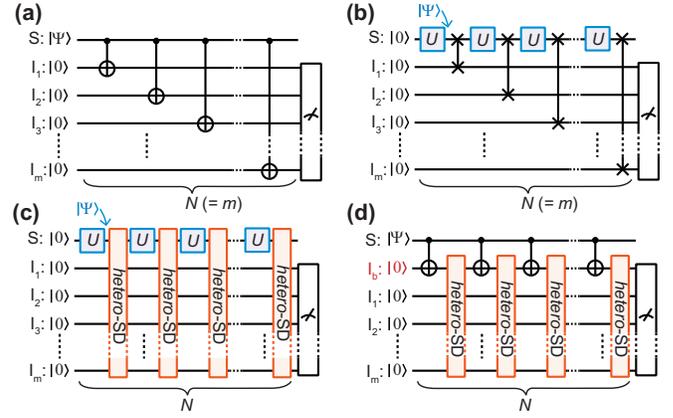}}
\caption{\label{fig1}
\small
(a) Spin amplification referred to by DiVincenzo~\cite{DiVincenzo00}.
(b) Spin amplification of the S spin response signal using selective SWAPs, and (c) that using the {\it hetero}-SD process.
(d) Scalable version of (a).
The I spins are initialized to $|0\rangle$ for simplicity, although they do not have to be in practice.
}
\end{figure}

In the S-I$^m$ spin system (a rarely-existing S spin and $m$ abundant I spins), the absolute maximum gain of spin amplification shown in Figs.~1(a) and 1(b) is limited by $m$.
To increase the maximum gain, $m$ must be increased.
Figs.~1(a) and 1(b) are, in principle, scalable, but they hardly scale up to larger $m$ in practice because they require individual addressing to all spins or a special spin network with tailored couplings such as a spin chain and a star topology~\cite{Perez,Kay,Lee07,JonesNOON}.
Until now, a gain as low as four has been demonstrated with nuclear spins in molecules in solution~\cite{Lee07,Cappellaro05}.
Building a spin amplifier with a high gain, say 100, seems as daunting as building a 100-qubit quantum computer.

In this Letter, we propose a new scalable implementation of spin amplification shown in Fig.~\ref{fig1}(c), which utilizes {\it spin diffusion}~\cite{SpinDiff} and {\it magnetic-field cycling}~\cite{NoackReview}.
The total Hamiltonian of an S-I$^m$ nuclear spin system with a dipolar coupling network, naturally existing in a bulk solid, is given by
\begin{equation}
\mathcal{H}=\omega_S S_Z + \omega_I \sum_i I_{iZ} + \sum_{i, j} \mathcal{H}_{i, j} + \sum_{i} \mathcal{H}_{i, {\rm S}},
\end{equation}
where $\omega_S$ and $\omega_I$ are the resonance frequencies of the rare spin S and the abundant spins I.
$\mathcal{H}_{i, j}$ and $\mathcal{H}_{i, {\rm S}}$ represent the homonuclear and the heteronuclear dipolar interactions, respectively.
The dipolar interaction, $\mathcal{H}_{i, j}$, between the $i$-th spin I$_i$ and the $j$-th spin I$_j$ (or S for heteronuclear case, $\mathcal{H}_{i, S}$) is given by
$
{\cal H}_{i, j}={\cal H}_{i, j}^{ZZ}+{\cal H}_{i, j}^{XY}
$
where
$
{\cal H}_{i, j}^{ZZ}=d_{i, j}I_{iZ}I_{jZ}
$
(for heteronuclear case, the dipolar coupling strength, $d_{i, j}$, is replaced with $d_{i, S}$) generates mutually controlled rotations about $z$ axis
and
$
{\cal H}_{i, j}^{XY}=-d_{i, j}(I_{iX}I_{jX}+I_{iY}I_{jY})/2
$
generates fractional power of iSWAP
.
Both are always on between homonuclear spins regardless of the static magnetic field strength.
Between heteronuclear spins, the former is always on and the latter is turned off in high static magnetic fields by the Zeeman energy difference being much larger than the dipolar interaction strength ($|\omega_S-\omega_I|\gg|d_{i, S}|$).
In low fields ($|\omega_S-\omega_I|\lesssim|d_{i, S}|$), the latter is turned on.
Under free time evolution of the total Hamiltonian in the high field, the $z$ components of only the I spins are shuffled by simultaneous fractional powers of iSWAPs, and this behavior is known as spin diffusion~\cite{SpinDiff}.
In the low field, the S spin is also involved in the spin diffusion among the I spins via its coupling to at least one I spin: we call it the heteronuclear-spin-involved spin diffusion ({\it hetero}-SD) process, where the total $z$ spin component is preserved.
The {\it hetero}-SD process can be switched on and off by magnetic-field cycling between the low and the high fields~\cite{NoackReview,Ivanov,19FMRI}.
Therefore the quantum circuit of Fig.~\ref{fig1}(c) can be easily realized by the simple repetition of synchronized field cycling and a radiofrequency (rf) pulse irradiation for the quantum operation $U$ to the S spin alone.
The process requires only a reasonably dense network of dipolar-coupled spins without requiring the tailored coupling strength ($d_{i, j}$), and therefore, is quite general, robust, and practical in real samples which may have some irregularities or randomness.

By exciting the $z$ component of the S spin with an rf pulse as $U$ and subsequently switching on the {\it hetero}-SD process for a sufficiently long duration, as shown in Fig.~1(c), the excited $z$ component of the S spin is diffused in the reservoir of the abundant I spins.
By repeating the sequence, the excitations are accumulated in the abundant I spins.
The information about whether the S spin is inverted by $U$ is converted into the decrease in the $z$ component of the total magnetization of $m$ abundant I spins, which is proportional to the I spin polarization.
Assuming that the polarizations of the S spin and the I spins are uniformly-mixed in the {\it hetero}-SD process, the I spin polarization after the $N$th repetition of the $U=$ NOT and the {\it hetero}-SD process decreases to $\epsilon_0[(m-1)/(m+1)]^N$ from the initial polarization $\epsilon_0$ of the S and the I spins.
Compared with the difference between the directly observed signal of the S spin of polarization $\epsilon_0$ and that of the inverted S spin, the expected gain of the signal obtained from the polarization difference of $m$ abundant I spins at the $N$th step of spin amplification with and without $U$ is given by
\begin{equation}
\label{eq:Gain}
G=\frac{m\left[\epsilon_0-\epsilon_0\left(\frac{m-1}{m+1}\right)^N\right]}{\epsilon_0-(-\epsilon_0)}=\frac{m}{2}\left[1-\left(\frac{m-1}{m+1}\right)^N\right].
\end{equation}

\begin{figure}[tbp]
\center{\includegraphics[width=.75\linewidth]{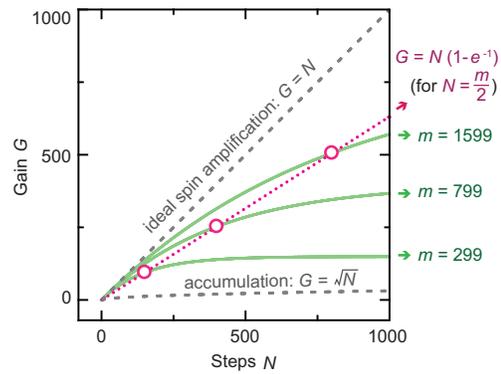}}
\caption{\label{fig2}
\small
The top dashed line shows the gain $G$ of the SNR of ideal spin amplification (Figs.~1(a) and 1(b)) with respect to the number of SWAP gates $N$.
The solid lines show the expected gain (Eq.~(2)) of the proposed one (Fig.~1(c)) with respect to the number of the {\it hetero}-SD process gates $N$ for different system sizes.
The bottom dashed line shows the gain of the direct simple repetitive detection with respect to the number of repetitions $N$.
}
\end{figure}

In an S-I$^m$ spin system, the gain of the spin amplification using the {\it hetero}-SD process increases with $N$ but is saturated to $m/2$ for $N \gg m/2$, as shown in Fig.~\ref{fig2}.
The proposed implementation using the {\it hetero}-SD process gives better SNR improvement with respect to $N$ than the simple $N$ repetitive detection.
When $m$ and $N$ are increased simultaneously while keeping $N=m/2$, the amplification gain scales up linearly as $m(1-e^{-1})/2$ (the dotted line) with respect to $m$ $(\gg 1)$.

The quantum circuit of Fig.~1(d) is a scalable version of Fig.~1(a) and could be used for a readout of quantum computers~\cite{footnote}.
While there is a functional difference between Figs.~1(d) and 1(c), 
the most difficult part, the scalability of implementation to achieve a higher gain, is common in Figs.~1(d) and 1(c).
In Fig.~1(d), the information about whether the state of a rare spin S is $|0\rangle$ or $|1\rangle$ is non-destructively copied to a buffer spin I$_b$ by the CNOT gate and accumulated in abundant I spins with {\it hetero}-SD processes.
The field cycling between the high field and the low field where the I-I$_b$ Zeeman energy difference is comparable to the I-I$_b$ coupling energy and the S-I$_b$ Zeeman energy difference is much larger than the S-I$_b$ coupling energy enables the {\it hetero}-SD process in which the I$_b$ spin is involved but the S spin is not.
Our implementations, both Figs.~1(c) and 1(d), using the field cycling do not require rf irradiation to the abundant I spins, of which a subtle imperfection causes the large degradation of the spin amplification gain.

\begin{figure}[tbp]
\center{\includegraphics[width=.96\linewidth]{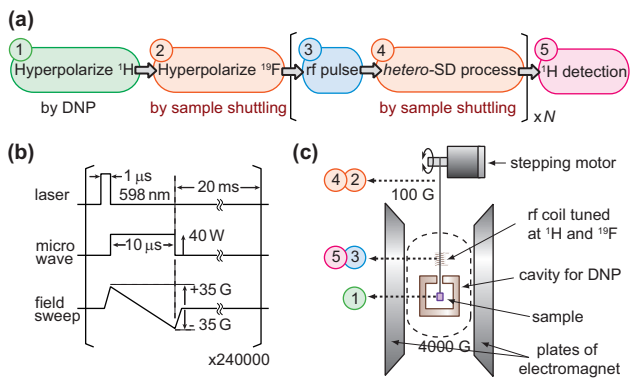}}
\caption{\label{fig3}
\small
(a) An experimental procedure of spin amplification using the {\it hetero}-SD process~\cite{Supp}.
(b) A DNP sequence with ISE~\cite{Henstra90}.
(c) A schematic diagram of an experimental setup where the sample position at each stage is indicated.
}
\end{figure}

In this work, we have performed experiments to amplify the response of a rare spin with respect to an rf pulse and demonstrated a spectroscopic application in an S-I$^{799}$ spin system in a bulk solid with the quantum circuit shown in Fig.~1(c).
We used a sample of a single crystal of naphthalene doubly-doped with $\sim$0.005~mol\% pentacene and $\sim$1~mol\% 2-fluoronaphthalene.
If the sample is uniform, it is approximately regarded as the ensemble of S-I$^{799}$ spin systems because a $^{19}$F spin is surrounded by 799 $^1$H spins on average.
The experimental procedure is shown in Fig.~\ref{fig3}(a)~\cite{Supp}.
In the first stage of every experiment, we enhanced the $^1$H spin polarization to $\sim$0.12 at 233~K by dynamic nuclear polarization (DNP) with photoexcited triplet electron spins of pentacene molecules~\cite{Henstra90,ViethReview,IinumaPRL,TakedaBook}, the sequence of the photoexcitation and the integrated solid effect (ISE)~\cite{Henstra90} shown in Fig.~\ref{fig3}(b).
The key stage in this experiment, the {\it hetero}-SD process, was realized by field cycling (4000~G $\to$ 100 G $\to$ 4000 G) using sample shuttling, as shown in Fig.~\ref{fig3}(c).
In 100 G, the resonance frequency difference between the $^1$H and the $^{19}$F spins is of the same order as the strength of the nearest neighbor $^1$H-$^{19}$F dipolar coupling.
After repeating an rf pulse to $^{19}$F spins as the quantum operation $U$ and the sample shuttling $N$ times, we observed the $^1$H magnetization.

We investigated the behavior of the $^1$H spin polarization with respect to the number of steps $N$ of spin amplification (Fig.~4(a)).
In the case of $U=$ {\bfseries\itshape I} with no rf pulse, the $^1$H spin polarization decayed with $\eta$ of $\sim$99.91\% per cycle, which gave close agreement with the decay rate calculated from $T_1$ measured at each sample position.
In the case of an on-resonance $\pi$ pulse (i.e. $U=$ NOT), the result was close to the theoretical decay $\epsilon_0\{(m-1)/(m+1)\}^N\eta^N$ (the dashed line) calculated from the decreased polarization in Eq.~(2) and the decay $\eta$.
The S spin response to $U$ is converted into the difference between the I spin polarization with and without $U$.
The polarization difference was amplified 37 times for $N=40$ and 136 times for $N=200$, compared with that for $N=1$.
The amplification gain anticipated from Eq.~(2) is 38 for $N=40$ and 157 for $N=200$.
By taking into account the frequencies for $^{19}$F and $^1$H spins, the signal obtained from the polarization difference of the $^1$H spins with and without $U=$ NOT was estimated to be $\sim$140 times as large as that of the $^{19}$F spin with the polarization 0.11.
It is noteworthy that the 140-time amplified signal is more than ten times as large as the signal of the perfectly polarized $^{19}$F spins.
Furthermore, a factor of 140 spin amplification in conjunction with the modern magnetic resonance force microscopy (MRFM~\cite{Rugar04}) technology, which is expected to detect a hundred nuclear spins in the near future~\cite{Mamin09}, may open the door to the detection of a faint signal from a single nuclear spin.

\begin{figure}[tbp]
\center{\includegraphics[width=.93\linewidth]{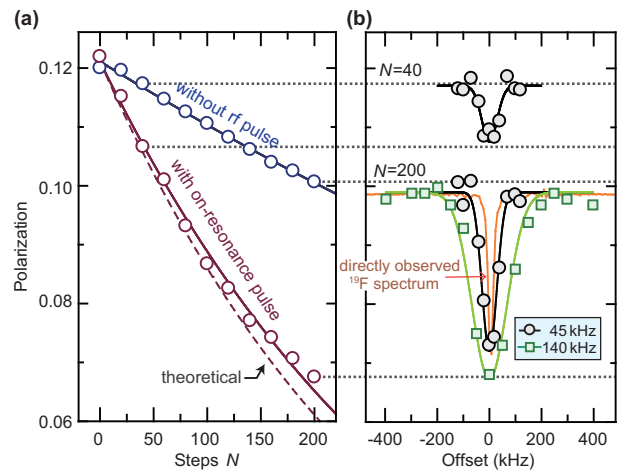}}
\caption{\label{fig4}
\small
(a) The behaviors of the $^1$H spin polarization with respect to the number of steps $N$ of spin amplification without rf pulse ($U=$ {\bfseries\itshape I} ) and with on-resonance $\pi$ pulses ($U=$ NOT).
The dashed line is the theoretical decay. 
(b) The circles show the spin-amplified frequency response spectra obtained for $N=40$ and 200.
They were obtained with the rf pulse with a peak amplitude of 45~kHz (the pulse length is $\sim$3 times as long as that of 140~kHz).
The squares show that for N=200 obtained with the rf pulse with a peak amplitude of 140~kHz.
Directly detected $^{19}$F spectrum vertically magnified is also shown for comparison.
}
\end{figure}


Amplified responses to the rf pulses with various carrier frequencies for $N=200$ give the excitation spectrum of the $^{19}$F spin, as represented by the squares in Fig.~\ref{fig4}(b).
The spin-amplified response spectra with the weaker rf pulse (the circles) were narrower than that but still broader than the directly observed $^{19}$F NMR spectrum.
The pulse intensity should be lowered in order to avoid power broadening.
The depth of the spectral response was increased with $N$.

All spin-amplified $^1$H NMR signals were reduced by a factor of a hundred on purpose to avoid saturation of the detection circuit~\cite{Supp}.
With the reduction, the directly observed $^{19}$F NMR signal was completely-concealed with the thermal noise.
The $^{19}$F signal observed without the reduction is shown in Fig. 4(b).
If the sample size is 100 times as small, the $^{19}$F signal can not be detected even without the reduction.
On the other hand, the spin amplified $^1$H signals of the small sample can be detected with the approximately same SNR as the present result without the reduction.
Spin amplification will have a distinct advantage over the simple repetitive detection in a smaller or a more diluted sample.
If we use a sample with a lower 2-fluoronaphthalene concentration, we could get a higher gain by increasing $N$ in the same manner.
The maximum achievable gain is inversely proportional to the concentration.
It has been reported that $T_1$ of the $^1$H spins in naphthalene doped with $\sim$0.01 mol\% pentacene in $\sim$7~G at 77~K is $\sim$166 minutes~\cite{IinumaPRL}, and the result implies that $T_1$ of a sample with a lower 2-fluoronaphthalene concentration at a lower temperature is longer.
In addition, there is room for speeding up field cycling~\cite{NoackReview,ReesePCCP}, and hence earning the number of steps $N$ well before the $^1$H polarization vanishes by T$_1$ relaxation.
Scalable spin amplification will enable practical applications of NMR spectroscopy to rarely-existing nuclear spins in infinitesimal bulk samples which have been concealed by thermal noise.
For higher sensitivity and spectral resolution, the quantum operation $U$ and the NMR detection should be performed in a higher magnetic field, which is realized by the field-cycling system described in~\cite{KagawaRSI}.

In the proposed implementation using field cycling for switching the {\it hetero}-SD process, the nuclear species are modestly restricted to those with the resonant frequency close to that of the abundant $^1$H spins.
For other nuclear species to be involved in spin diffusion process, the double resonance irradiation satisfying the Hartmann-Hahn condition~\cite{HH}, as in the cross polarization experiment~\cite{PinesCP} may be used, although it is subject to the rf pulse imperfection and the much faster decay with $T_{1\rho}$.
State-of-the-art high fidelity pulse engineering~\cite{KhanejaGRAPE,Ryan08} may be required to implement high-gain spin amplification.

The demonstrated spin amplification (Fig.~1(c)) can be straightforwardly modified to Fig.~\ref{fig1}(d).
It can be realized, for example, by doping 2-$^{13}$C 2-fluoronaphthalene, in which the $^{13}$C, $^{19}$F, and $^1$H spins play the roles of the S, the I$_b$, and the I spins, respectively.
Recently, optically detected magnetic resonance (ODMR)~\cite{Wrachtrup93,Kohler93} has accomplished single-event state-readout of a single nuclear spin~\cite{Neumann10}, which is attracting attention in quantum information science.
The breakthrough of the gain shown in this Letter is an important step for state-readout of a single nuclear spin in bulk solids with more versatile detection such as MRFM and the inductive detection.

In this Letter, we have proposed a scalable and practical implementation of spin amplification and successfully demonstrated a gain as large as 140.
A faint magnetization of a rare spin has been quantum-logically transferred beyond the molecular boundary to abundant spins in the bulk solid sample through spin diffusion process which does not require individual addressing nor a specially tailored spin network.
The amplified signal obtained from the abundant spins is more than ten times as large as that of the completely polarized rare spins. 
A spectroscopic application has been demonstrated by 140-time amplified frequency response spectra of the rare spin.
It will enable practical applications of NMR spectroscopy to infinitesimal samples which have been concealed by thermal noise. 
A higher gain is possible in the sample with a lower concentration of rare spins.
Spin amplification scalable to a higher gain opens the door to a single spin measurement and a readout of spin-based quantum computers.

\begin{acknowledgments}
We thank Kazuyuki Takeda for fruitful discussions.
This work was supported by the CREST program of JST, MEXT Grant-in-Aid for Scientific Research on Innovative Areas 21102004, and the Funding Program for World-Leading Innovative R\&D on Science and Technology (FIRST).
M.N. and K.T. were also supported by the Global-COE Program of Osaka University.
\end{acknowledgments}


\begin{thebibliography}{10}
\bibitem{footnote}
Fig.~1(d) is similar to the last scheme in Ref.~\cite{Cappellaro05} but is simpler than that.
The latter relies on multiple pulse sequences to establish entanglement while the former does not require any rf irradiation to the abundant I spins.
\end{thebibliography}

\section*{Supplemental Material}

Naphthalene was extensively purified by the zone melting method and then a single crystal of naphthalene doubly-doped with $\sim$0.005~mol\% pentacene and $\sim$1~mol\% 2-fluoronaphthalene was grown by the Bridgman method.
It was cut into a piece with a weight of $\sim$2.5 mg and mounted in a cylindrical cavity (TE011, 12.08 GHz) so that the long axis of the pentacene molecules was parallel to the static magnetic field~\cite{sAbrahams,sStrien}.
Supplementary Figure~1 is a detailed drawing of the experimental setup.

DNP with photoexcited triplet electron spins was performed in the static field of 4000 G.
We show the energy diagram of pentacene in Supplementary Figure~2.
598 nm laser irradiation was applied to the sample, so that the pentacene molecules underwent intersystem crossing from the excited singlet state to the excited triplet state.
The population distribution over the triplet sublevels was highly biased in this orientation~\cite{sSloop}.
At the moment, the high non-equilibrium electron spin polarization was transferred to the $^1$H spins in pentacene molecules by ISE~\cite{sHenstra90}.
In the ISE, the microwave irradiation and the field sweep were applied in such a way that the inhomogeneously-broadened electron spin packets were adiabatically swept over, and the Hartmann-Hahn condition was satisfied between the electron spins in the rotating frame and the $^1$H spins in the laboratory frame at a certain moment during the adiabatic passage.
The electrons were returned to the ground state immediately after the ISE.
We accumulated the $^1$H spin polarization by repeating the photoexcitation and the ISE at sufficiently long intervals, during which the $^1$H polarization localized in pentacene diffused to $^1$H spins in the sample.
It has been reported that the $^1$H polarization in a single crystal of 0.018 mol\% pentacene doped naphthalene has been enhanced to $\sim$0.7 in 3000~G at 105 K~\cite{sTakedaJPSJ}.

The sample shuttling from the center of the cavity in 4000~G to 13.5 centimeters above it in 100~G took $\sim$0.67 seconds, and then, the sample stayed for 10 milliseconds and returned to an rf coil in 4000~G in $\sim$0.67 seconds.
In such a low field as 100 G, polarization transfer occurs even between heteronuclear spins.
This spin dynamics mechanism has been studied for transferring {\it para}-hydrogen induced polarization (PHIP) into a molecule in solution~\cite{sIvanov,s19FMRI}.
After enhancing the $^1$H polarization to $\sim$0.12 by DNP and sample shuttling, the $^{19}$F spin polarization measured by $\pi/2$ pulse excitation was $\sim$0.11, and therefore, the transfer efficiency of the polarization was $\sim$92\%.
The whole sample path was thermally insulated with a double-layered glass tube and cooled with nitrogen gas flow down to 233~K in order to suppress naphthalene sublimation and reduce T$_1$ relaxation.
$T_1$ of the $^1$H spins in 4000~G was $\sim$212 minutes and that in 100~G, which was the lowest field on the sample path, was $\sim$34 minutes at 233 K.
Therefore, the total magnetization was scarcely decreased by the sample shuttling.
The present sample system, which has very long $T_1$ and can be hyperpolarized, is a good test-bed for spin amplification experiments.

We used the Hermite 180$^\circ$ pulse~\cite{sWarren84} with a peak amplitude of 140~kHz shown in Supplementary Figure~3a as the quantum operation $U$.
As shown in Supplementary Figure~3b, its excitation range is broader than the inhomogeneous broadening of the $^{19}$F spin in the sample, and the decrease in magnetization with the pulse at the offset of more than 300 kHz is less than 0.1\%.
We also used the Hermite 180$^\circ$ pulse with a peak amplitude of 45~kHz, of which the pulse length was $\sim$3 times as long as that of 140~kHz.

The $^1$H NMR signal was excited by a 5$^\circ$ pulse and attenuated with -20~dB and, therefore, reduced by a factor of more than a hundred on purpose to avoid saturation of the detection circuit.
The $^1$H and $^{19}$F polarizations were determined by comparing them with the $^1$H signal of ethanol and the $^{19}$F signal of 2,3,4-trifluorobenz-aldehyde, respectively, in thermal equilibrium at 4000~G and 233~K.
The polarization is defined by $(N_{|\uparrow\rangle}-N_{|\downarrow\rangle})/(N_{|\uparrow\rangle}+N_{|\downarrow\rangle})$, where $N_{|\uparrow\rangle}$ and $N_{|\downarrow\rangle}$ are the populations of up and down spins.
The pulse shape, NMR detection, DNP sequence, and motor control were programmed by a homebuilt spectrometer~\cite{sFPGA,sOPENCORE}.

\begin{figure}[htbp]
\center{\includegraphics[width=.6\linewidth]{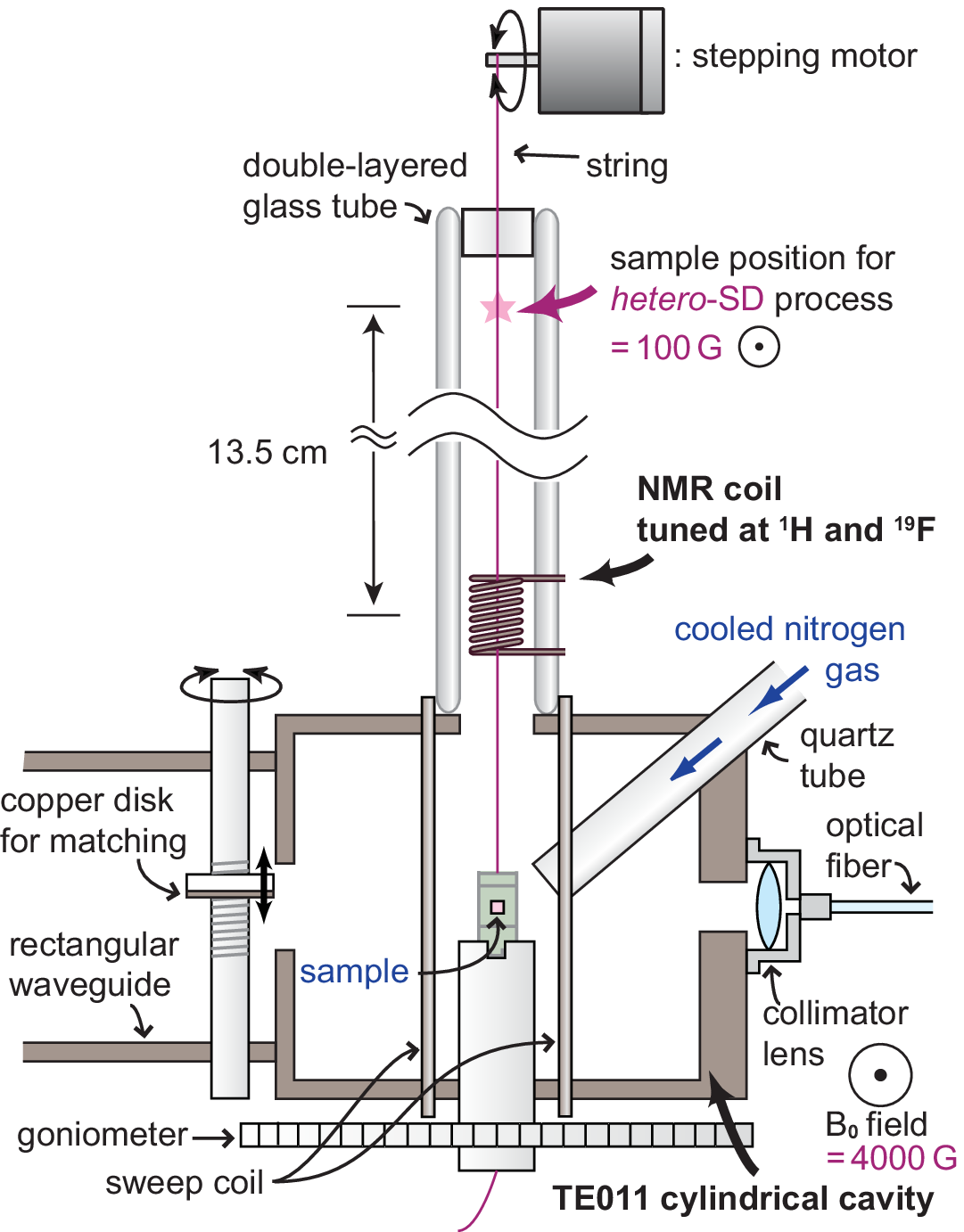}}
\end{figure}
\noindent Supplementary Figure 1.
{\bf Detailed drawing of the experimental setup.}

\begin{figure}[htbp]
\center{\includegraphics[width=.4\linewidth]{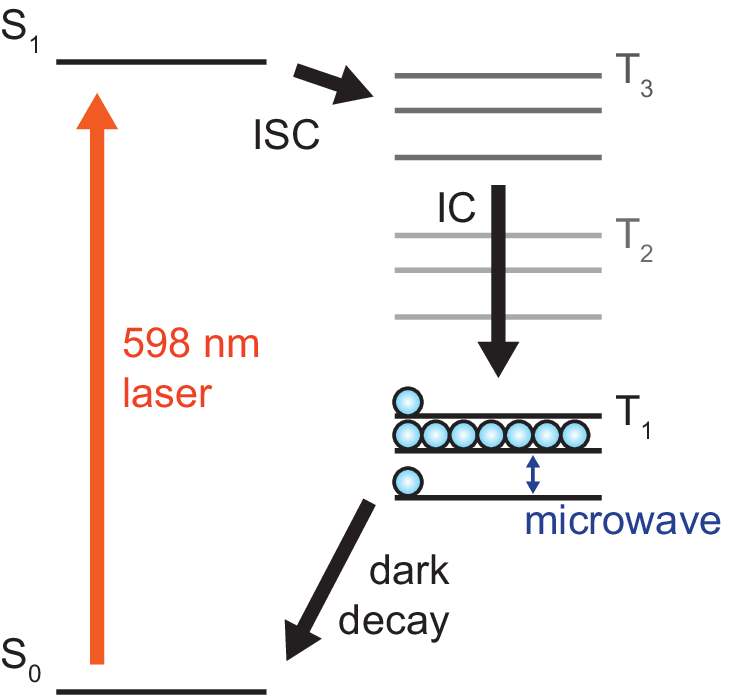}}
\end{figure}
\noindent Supplementary Figure 2.
{\bf Energy diagram of pentacene.}
ISC and IC mean intersystem crossing and internal conversion, respectively.

\begin{figure}[htbp]
\center{\includegraphics[width=\linewidth]{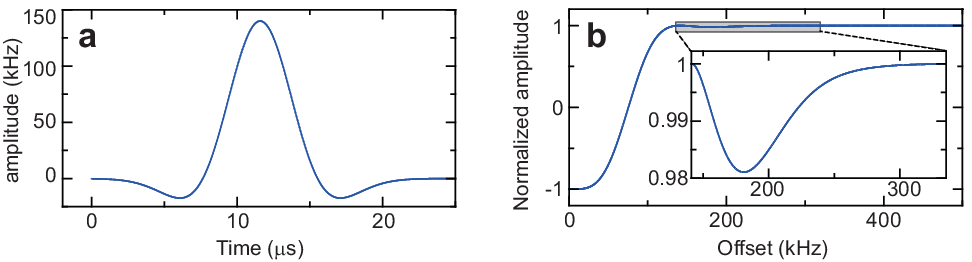}}
\end{figure}
\noindent Supplementary Figure 3.
{\bf Hermite 180$^\circ$ pulse.}
{\bf a}, The amplitude profile of the Hermite 180$^\circ$ pulse of the peak amplitude 140~kHz.
{\bf b}, The excitation frequency response to this Hermite 180$^\circ$ pulse with respect to the carrier frequency offset.
The vertical axis shows the remaining longitudinal magnetization after the pulse, which is normalized with the initial magnetization.

\clearpage
\end{document}